\documentclass[12pt]{article}
\usepackage[centertags]{amsmath}
\usepackage{amsfonts}
\usepackage{bm}
\begin{document}
\title{\textbf{
Quantization of a torus  phase space}} 
\author{H.S.Sharatchandra\thanks{E-mail:
{sharat@cpres.in}
} \\[2mm]
{\em Centre for Promotion of Research,} \\
{\em 7, Shaktinagar Main Road, Porur, Chennai 600116, India}
}
\date{}
\maketitle
\begin{abstract}
Quantization of $R^2$ and $S^1 \times S^1$ phase spaces are explicitly carried out tweaking the
techniques of geometric quantization. Crucial is a combined use of left and
right invariant vector fields. Canonical  bases, operators and their
actions are explicitly presented. Arguments of Dirac and also Wu and Yang for monopoles
are applied for obtaining the quantization of the phase space area. 
Equivalence of  states in the infinite dimensional prequantum Hilbert space resulting in a  physical Hilbert space of 
dimension equal to the phase space 
area in units of the Planck constant is demonstrated. These techniques can be applied to any manifold with a
symplectic structure.

\end{abstract}

\smallskip

\smallskip


\section{Introduction}\label{i}
In this paper we carry out quantization of $R^2$ and $S^1 \times S^1$  phase spaces
tweaking the techniques
geometric quantization \cite{w}. We advocate 
a combined use of left and right invariant vector fields 
on the  Heisenberg-Weyl group. Our aim is an explicit presentation of canonical bases, operators and 
their actions. We use the arguments of Dirac \cite{d} and also Wu and Yang \cite{wy} for monopoles
to get quantization of the phase space  area. In the big prequantum Hilbert space we obtain eqivalences
of wave functions, in the sense they  
have same physical matrix elements. This results in a physical Hilbert space that is equivalent to that of
the Schr\"{o}dinger formalism. In case of a finite phase space area as for the torus, the 
physical Hilbert space has exactly 
the dimension of the phase space  area in units of the Planck constant.

Each step of our technique can be carried out for any classical phase space with a symplectic structure \cite{sg}.

There is extensive work on quantum cat map on a torus, especially for understanding quantum chaos
\cite{hb}. This analyzes quantum time evolution on a torus as a phase space. Periodicity arguments are 
used to get quantization of the phase space area.  The position or momentum basis states are (periodic)
delta functions. Our approach and arguments are different. We use phase space wave functions.
 
\section{Quantum mechanics on a $R^2$ phase space}\label{psqm}
The position and momentum operators $\overleftarrow{Q},\overleftarrow{P}$ in quantum mechanics satisfy the Heisenberg 
commutation  relation
\begin{equation}\label{cr}
[\overleftarrow{Q},\overleftarrow{P}]=i\hbar.
\end{equation}
We choose 
\begin{eqnarray}\nonumber
\overleftarrow{Q} &=& q+i\hbar\frac{\partial}{\partial p}\ ,
  \\\label{I}
\overleftarrow{P}&=&-i\hbar\frac{\partial}{\partial
q}\ .
\end{eqnarray}
These are supposed to act on wave functions $\psi(q,p)$ belonging to  Hilbert space
 $L^2( \mathbb{R}^2)$.
This choice  has a long history. It can be related to  vector field that is invariant under  
the left action of the Heisenberg-Weyl group $H_3$.
The Weyl or displacement operator is
\begin{equation}\label{do}
  D(q,p)  = e^{i(p\overleftarrow{Q} - q\overleftarrow{P})/\hbar}\ .
\end{equation}
Using the well-known relation
\begin{equation}\label{lg}
 e^{\hat A}e^{\hat B}=e^{{\hat A}+{\hat B}+[{\hat A},{\hat B}]/2},
\end{equation}
which is valid when operator $[{\hat A},{\hat B}]$ commutes
with both $\hat A$ and $\hat  B$ we get the multiplication rule
\begin{equation}\label{hwg}
  D(b,a)   D(q,p)  = D(q+b,p+a)e^{i(aq-bp)/(2\hbar)}\ .
\end{equation}

Crucial for us is the existence of another set of operators ${\overrightarrow Q},{\overrightarrow P}$ on $\psi(q,p)$ also satisfying 
Heisenberg commutation  relation and commuting with ${\overleftarrow Q},{\overleftarrow P}$.
 \begin{eqnarray}\nonumber
\overrightarrow{Q} &=& i\hbar\frac{\partial}{\partial p}\ ,
  \\\label{II}
\overrightarrow{P}&=&p+i\hbar\frac{\partial}{\partial
q}\ ,
\end{eqnarray}
satisfy  
\begin{equation}\label{pcr}
[\overrightarrow{Q},\overrightarrow{P}]=i\hbar.
\end{equation}
This is related to the vector field that is  invariant under the right action of the  Heisenberg-Weyl group.
As the multiplications on left and right commute,
this pair commutes with each of $\overleftarrow{Q},\overleftarrow{P}$. 
\begin{equation}\label{pcr1}
[\overleftarrow{Q},\overrightarrow{P}]=0, [\overrightarrow{Q},\overleftarrow{P}]=0,
[\overleftarrow{Q},\overrightarrow{Q}]=0, [\overrightarrow{P},\overleftarrow{P}]=0.
\end{equation}
For a recent use of these operators see for eg., \cite{mjt},  \cite{sm}.
The operators  $\overleftarrow{Q},\overleftarrow{P}$ are the 
"physical operators"
 in the sense  that we are interested in expectation values of operators built from them.
It appears that the  operators  $\overrightarrow{Q},\overrightarrow{P}$ are unphysical in the sense  
that they are incidental to the 
formalism and we are not interested in their expectation values. We shall refer to them as the "shadow
operators".

For a (non-normalizable) complete set of basis we  choose simultaneous  eigenstates of
a set of commuting operators. If we choose 
 the 
commuting set $\overrightarrow{Q}$ and $\overleftarrow{P}$ we get the conventional  plane waves
\begin{equation}\label{bs1}
\phi_{lk}(q,p) = e^{i(kq-lp)/\hbar}, 
\end{equation}
where labels $k$, $l$ are the eigenvalues of $\overleftarrow{P}$ and $\overrightarrow{Q}$ respectively.
We shall refer to this basis as the P-basis, as the physical operator $\overleftarrow{P}$ is diagonal.

It is more interesting to choose the 
commuting set $\overleftarrow{Q}$ and $\overrightarrow{P}$. 
\begin{equation}\label{bs2}
\psi_{lk}(q,p) =e^{ipq/\hbar} e^{-i(kq+lp)/\hbar}, 
\end{equation}
where labels $k$, $l$ are the eigenvalues,
\begin{eqnarray}\nonumber
\overrightarrow{P}\psi_{lk}(q,p) =k \psi_{lk}(q,p), 
\\\label{obs2}
\overleftarrow{Q}\psi_{lk}(q,p)=l \psi_{lk}(q,p).
\end{eqnarray}
We shall refer to this basis as the Q-basis, as the physical operator $\overleftarrow{Q}$ is diagonal.

We now explain the unusual phase  factor $e^{ipq/\hbar}$. We relate the choice (\ref{I}) to 
covariant derivatives \cite{w}  in the presence of a $U(1)$ gauge field,
\begin{equation}\label{cd}
D_i={\partial}_i -iA_i,
\end{equation}
 with
\begin{equation}\label{pot}
A_p=\frac{1}{\hbar}q, A_q=0,
\end{equation} 
 as the gauge potential: 
\begin{equation}\label{D}
\overleftarrow{Q}=i\hbar D_p,\overleftarrow{P}=-i\hbar D_q.
\end{equation} 
The corresponding
"magnetic field" $ B= {\partial}_q A_p-{\partial}_p A_q$ is a constant independent of $q$ and $p$, 
\begin{equation}\label{mf}
B=\frac{1}{\hbar}.
\end{equation}
We may cancel the effect of the gauge field using  a path dependent (i.e. a non-integrable) phase factor 
$e^ {i\int _C d\xi_i A_i(\xi)}$. For path C we choose a standard path starting from
the  origin $(0,0)$ along the q-axis to the point $(0,p)$ and  then parallel to the p-axis to
$(q,p)$. This gives the  phase factor denoted by the  first exponential in  (\ref{bs2}).
Apart from this
we may choose basis states of $i{\partial}_i$  i.e. plane wave states as denoted by the second exponential in  (\ref{bs2}).

The inner product of two states $ \psi_1$,$ \psi_2$ is defined to be 
\begin{equation}\label{ip}
\langle \psi_1|\psi_2\rangle =  \int \psi_1^*(q,p)\,
    \psi_ 2(q,p)\, \frac{dqdp}{h}\ .
\end{equation}
Note that the integration is over both $q$ and $p$, i.e. the phase space. This gives the prequantum Hilbert space
\cite{w}. It is "too big" compared to the Hilbert space of the Schr\"{o}dinger formalism, as the wave functions are functions of both the coordinate and the momentum variables. We will show equivalences among wave functions of this space. 

In spite of the unusual factor $exp(ipq/\hbar)$ in the basis functions (\ref{bs2}), we have the 
Dirac delta function orthonormality,
\begin{equation}\label{on}
\langle \psi_{lk}|\psi_{l'k'}\rangle =h \delta(l-l')  \delta(k-k').
 \end{equation}

As is to be expected the basis functions (\ref{bs2}) are not eigenfunctions of  $\overleftarrow{P},\overrightarrow{Q}$:
\begin{eqnarray}\nonumber
\overleftarrow{P}\psi_{lk}(q,p) =(p-k) \psi_{lk}(q,p),
\\\label{obs22}
\overrightarrow{Q}\psi_{lk}(q,p)=(-q+l) \psi_{lk}(q,p).
\end{eqnarray}
It  is more enlightening to rewrite the eigenfunctions as
\begin{equation}\label{pbs2}
\psi'_{lk}(q,p) =e^{i(p-k)(q-l)/\hbar},  
\end{equation}
which differs from  (\ref{bs2}) by ($k$, $l$ dependent) phase factors.
Now we see the actions of 
$e^{ia\overrightarrow{Q}/\hbar}$ and $e^{-ib\overleftarrow{P}/\hbar}$ as shifting the eigenvalues $k$, $l$  by $a$ and 
$b$ respectively:
\begin{equation}\label{obs21}
e^{ia\overrightarrow{Q}/\hbar}\psi'_{lk}(q,p) =\psi'_{l,k+a}(q,p), 
\end{equation}
 \begin{equation}\label{obs22}
e^{-ib\overleftarrow{P}/\hbar}\psi'_{lk}(q,p)=\psi'_{l+b,k}(q,p).
\end{equation}

A general state of the Hilbert state is a linear combination of the basis states (\ref{bs1}) or (\ref{bs2}). 
In spite of the unusual appearance of (\ref{bs2}), we may formally expand the basis states (\ref{bs2}) in the basis (\ref{bs1})
\begin{equation}\label{bse}
\psi_{lk}(q,p) =\int 
    \phi_{l'k'} (q,p)\, e^{ipq/\hbar} e^{-i(k+k')(l-l')/\hbar}\frac{dl'dk'}{h}. 
\end{equation}
Strictly speaking this is to be viewed as relating the coefficients of expansion of a nomalizable state in the two bases.
 
Note that the physical operators $\overleftarrow{Q},\overleftarrow{P}$ are diagonal in the lable $k$ 
in the Q-basis and in addition their actions are insensitive to it (\ref{obs21},\ref{obs22}). As a consequence, the 
variable $q$ in (\ref{bs2}) drops out in calculation of the matrix elements.This is the way the 
variable $q$ is redundant for the 
physical measurements and phase space quantum mechanics gets  related to the Schr\"{o}dinger formalism.  

Consider any unitary operator $U(\overrightarrow{Q},\overrightarrow{P})$ built  using the shadow operators
$\overrightarrow{Q}$, and $\overrightarrow{P}$. When acting on  states $\psi_{lk}(q,p)$ this gives a states
which have eactly same  matrix elements as in its absence, for any physical observable built out of the physical  operators
 $\overleftarrow{Q},\overleftarrow{P}$. This is to be regarded as a huge gauge invariance. This can be handled 
by various techniques used in gauge theories \cite{sg}.

\section{Quantum mechanics on a torus phase space}\label{t}

We now consider a finite phase space. We consider a torus of size $a$ in variable $p$ and size $b$ in variable $q$.
We represent the torus by a
rectangle with the coordinates $(q,p)$ of the corners at
\begin{equation}\label{r}
(0,0),(b,0),(0,a), (b,a),
\end{equation} 
with opposite edges identified.

The commuting operators  $\overrightarrow{Q}$ and $\overleftarrow{P}$ look innocuous 
and the requirement of  periodic boundary conditions would give the basis functions
\begin{equation}\label{tbs1}
\phi_{nm}(q,p) =exp\Big( 2\pi i\big(\frac{mq}{b}-\frac{np}{a}\big)\Big),
\end{equation}
$ m,n= 0,\pm 1,\pm2,\cdots$. At this stage the Hilbert space is infinite 
dimensional. This is the P-basis in which $\overleftarrow{P}$ has the eigenvalues $mh/b$.

On the other hand if we consider the commuting operators  $\overleftarrow{Q}$ and $\overrightarrow{P}$
we have the Q- basis functions
\begin{equation}\label{tbs2}
\psi_{nm}(q,p) =exp\Big( 2\pi i\big(\frac{pq}{h}-\frac{mq}{b}-\frac{np}{a}\big)\Big).
\end{equation}
This appears to have serious problems. Due to the unusual phase factor $exp\big(2\pi ipq/h\big)$   the
 periodic 
boundary conditions cannot be satisfied.

The resolution to this problem lies in Dirac's treatment \cite{d} of the quantum mechanics of a charged 
particle in the
field of a magnetic monopole. Dirac argued that a quantum mechanical  wave function can have 
a non-integrable phase factor. There is
a strong relation to the present problem. We now have  a "charged particle" in the presence of 
a uniform magnetic field   $1/\hbar$
on a torus (See \ref{pot}). This is possible only if we imagine that a  "magnetic  monopole" of charge $area/\hbar$ is present 
inside the torus producing a magnetic field $1/\hbar$ on the torus.For example there can be a loop of 
"magnetic  monopole" of line  density 
$b/ \hbar$  lying inside the torus. 
We may imagine a sheet of magnetic flux $ab/\hbar$  entering into the torus along the edge $q=0 	~(mod ~b)$
 which then spreads 
out to give a uniform flux on the torus.This sheet is the analogue of the Dirac string.
The  gauge potential (\ref{pot}) is not periodic.
 We may interpret 
it to have
singularity at  $q=0 ~(mod ~b)$ where the Dirac sheet enters the torus. 

In his unique way Dirac argued that the consistency of the quantum system requires that  the monopole 
charge be quantized, in 
which case  the singular string is invisible. Similar arguments are applicable to the present case with striking 
conclusions. Consider 
a loop close to the edge of the rectangle (\ref{r}). This corresponds to a loop that goes in opposite directions
at either sides of the edges $p=0 (mod ~a)$ and $q=0 ~(mod ~b)$. Due to the discontinuity in the gauge 
potential at $q=0 ~(mod ~b)$
the charged particle acquires an extra phase $exp (iab/\hbar)$. The  singular sheet is invisible if
{\bf area/h is an integer N}. Thus
{\it quantum theory on a phase space area $A$  is  consistent only if the area is an integral multiple of the
Planck constant}. In this case phase factor at  $q=b$ is $exp(2\pi i N p/a)$, so that periodicity in $p$ is restored. 
In literature on geometric quantization \cite{w} this is referred to as the Bohr-Sommerfeld condition and is related to the 
first Chern class of the line bundle.

It is worth our while to formulate the problem in the way of Wu and Yang  \cite{wy} using the language of 
vector bundles.We use two overlapping coordinate charts to cover the torus:
$I:-\delta <q < \pi+\delta, II:\pi-\delta <q < 2\pi+\delta$ for all $p$. The wave functions are chosen as in
(\ref{tbs2}) in both these regions. They match in the overlap $\pi-\delta <q < \pi+\delta$. But in the other
 overlap they differ in  phase:
\begin{equation}\label{gt}
 -\delta <q < \delta:
\psi_{II}(q,p) =e^{ i bp/\hbar} \psi_{I}(q,p).
\end{equation}
This phase can be regarded as a gauge transformation to match  the wave functions of the two regions in this
overlap. This is periodic in $p$ only if $ab=Nh$.

The inner product  is defined to be 
\begin{equation}\label{tip}
\langle \psi_1|\psi_2\rangle =  \int\limits_{0}^{a} \int\limits_{0}^{b}\psi_1^*(q,p)\,
    \psi_ 2(q,p)\, \frac{dq}{b}\frac{dp}{a}\ .
\end{equation}

Note that the range of integration over $q$
and $p$ is finite.Indeed $\psi^{*}_1 \psi_2$ should be computed in regions $I$ and $II$ separately,
but as this is gauge invariant and also periodic, integration can be carried out in the range $0<q<b,0<p<a $ directly. 
The basis functions (\ref{tbs2}) are orthogonal and normalized. 
Note that due to the area quantization the inner product is
\begin{equation}\label{tip1}
\langle \psi_1|\psi_2\rangle =  \int \limits_{0}^{a} \int\limits_{0}^{b}\psi_1^*(q,p)\,
    \psi_ 2(q,p)\, \frac{dq dp}{Nh}\ .
\end{equation}

In Schr\"{o}dinger formalism with a   periodic coordinate variable   (for example, a  planar rotor)
the coordinate operator $\overleftarrow{Q}$  has no meaning because 
it is not periodic. Only 
operators such as $e^{2\pi i\overleftarrow{Q}/b}$ are meaningful.Relating the coordinate operator 
$\overleftarrow{Q}$ to a 
covariant derivative (\ref{D}) has the  great advantage of making  it directly meaningful.
Our basis states $\psi_{nm}(q,p)$ are eigenstates of the 
$\overleftarrow{Q}$ with eigenvalues $nh/a$. 

As with (\ref{pbs2}) it is revealing to write the Q-basis as
\begin{equation}\label{tpbs2}
\psi'_{nm}(q,p) = e^{2\pi i N(\frac{p}{a}-\frac{m}{N})(\frac{q}{b}-\frac{n}{N})}.
\end{equation}
We see immediately,
\begin{equation}\label{tobs21}
e^{-2\pi i\overleftarrow{P}/a} \psi'_{nm}(q,p) =\psi'_{n+1, m}(q,p),
\end{equation}
\begin{equation}\label{tobs22}
e^{2\pi i\overrightarrow{Q}/b} \psi'_{nm}(q,p) =\psi'_{n, m+1}(q,p).
\end{equation}
Thus the  operator $e^{-2\pi i\overleftarrow{P}/a}$  shifts $n$ by one.  
It as also its (positive or negative) powers are  valid operators  in the Hilbert space. Similarily
$e^{2\pi i\overrightarrow{Q}/b}$ shifts $m$ by one.

\section{One state per Planck constant phase space area}\label{h}
The physical operators built with $\overleftarrow{Q}$ and $\overleftarrow{P}$ are diagonal in $m$. States with same
$n$ but different $m$ are equivalent. Therefore we may set (say)
 $m=0$ for any matrix elements involving  "physical" operators.
Even after this the Hilbert space has an infinite dimension as $n$ takes all positive or negative integer values.
It appears that by taking a Fourier series of the basis
states (\ref{tbs1}) we can construct an eigenstate of  $\overleftarrow{P}$ that is a (periodic) Dirac delta function. But we
should  expect that with a phase space of a finite area, momentum (or position) cannot be localized indefinitely.

Heisenberg commutation  relations gives
\begin{equation}\label{hw2}
e^{2\pi i\overleftarrow{Q}/b} e^{-2\pi iN\overleftarrow{P}/a}=e^{-2\pi iN\overleftarrow{P}/a} e^{2\pi i\overleftarrow{Q}/b} 
 e^{2\pi iNh/(ab)}.
\end{equation}
Precisely because of our quantization of area the last exponential is unity. Therefore we  can consistently
impose an additional equivalence of states given by the operator equivalences,

\begin{equation}\label{oeq1}
e^{-2\pi iN\overleftarrow{P}/a} \equiv 1,
\end{equation}
\begin{equation}\label{oeq2}
e^{2\pi iN\overrightarrow{Q}/b} \equiv 1.
\end{equation}
This makes states $\psi'_{nm}(q,p)$ shifted in $n$ (or $m$) by $N$ equivalent:
\begin{eqnarray}\nonumber
\psi'_{n+N,m}(q,p) \equiv \psi'_{nm}(q,p),\\\label{eq1}
\psi'_{n,m+N}(q,p) \equiv \psi'_{nm}(q,p).
\end{eqnarray}
Note that   in the basis states (\ref{tpbs2})  shifting $m$ to $m+N$ can be absorbed by increasing $p$ to $p+a$ .
By the proposed equivalence we are recovering  periodicity in $p$ in the basis states. In the same way
periodicity in $q$  is also realized.

With these  equivalences, distinct basis states can be labeled by a finite set of indices, $m,n=0,1,...,(N-1)$. Further,
to calculate  any physical observable it suffices to set say $m=0$. This way the dimension of the "physical"
Hilbert space is reduced to $N$. This corresponds to exactly one state per area $h$ of the phase space. 

We can draw an analogy with a crystal lattice. Consider $N$ lattice points spaced equally on a circle of circumference
$b$. The normal modes are   $exp(2\pi inq/b)$ with $n$ taking $N$ values $n= 0,1,...,(N-1)$.
The mode with some $n$ shifted by a multiple of $N$ produces same amplitudes at the $N$ lattice points and therefore
physically indistinguishable.This is the way equivalence of modes shifted by $N$ works, giving a finite dimensional
Hilbert space. Strictly speaking the wave vector is meaningful only modulo $2\pi N/b$.

With these  equivalences $\overleftarrow{P}$ (and $\overrightarrow{Q}$) are no longer good operators on our states. 
This should be expected.
The Heisenberg commutation  relations cannot be realized on a finite dimensional Hilbert space, as the trace of the 
commutator is zero whereas trace of the identity operator is the dimension of the  Hilbert space. However 
$e^{-2\pi i\overleftarrow{P}/a}$ and $e^{2\pi i\overrightarrow{Q}/b}$ and their (positive or negative) powers are good operators.

For our basis states (\ref{tpbs2}) the following relations also hold as identities:
\begin{equation}\label{i2}
e^{-2\pi iN\overrightarrow{P}/a}=1,
\end{equation}
\begin{equation}\label{i1}
e^{2\pi iN\overleftarrow{Q}/b}=1.
\end{equation}
as they are eigenstates of $\overleftarrow{Q}$ and $\overrightarrow{P}$ with eigenvalues $nb/N$ and $ma/N$ respectively.

If we started with  $\overleftarrow{P}$ and $\overrightarrow{Q}$ we would have naturally chosen the P-basis states (\ref{tbs1}). They are
also as good a choice. Now the role of the equivalences (\ref{oeq1},\ref{oeq2}) and the identities (\ref{i1},\ref{i2}) get interchanged. Now we
have eigenstates of these operators and the other set, $\overleftarrow{Q}$ and $\overrightarrow{P}$  serve as
shift  operators on the  Hilbert space. This way both choices are consistent, the choice depending on which 
physical operator we want to diagonalize.

We shall formally denote the two sets of basis functions after taking into account all the equivalences by
$\Phi_{n,m}, n,m=0,1,2,\ldots ,(N-1)$ for the P-basis and $\Psi_{nm}, n,m=0,1,2,\ldots ,(N-1)$ for the Q-basis. 
In Table 1 we  explicitly present  the action of the relevant operators on them. From these actions we can
also read off the expansion of the  Q-basis in the P-basis.

\begin{table}[!t]
\centering
\renewcommand{\arraystretch}{1.2}
\begin{tabular}{|c|c|c|}
\hline
$Operator$& $\Phi_{nm}$&$\Psi_{nm}$ \\
\hline
$exp(-2\pi i\overleftarrow{P}/a) $& $e^{ -2\pi im/N}\Phi_{n,m}$&$\Psi_{n+1,m}$ \\
\hline
$exp(2\pi i\overleftarrow{Q}/b)$& $\Phi_{n,m+1}$&$e^{2\pi i n/N}\Psi_{n,m}$ \\
\hline
$exp(-2\pi i\overrightarrow{P}/a)$& $\Phi_{n+1,m}$&$e^{-2\pi i m/N}\Psi_{n,m}$ \\
\hline
$exp(2\pi i\overrightarrow{Q}/b)$& $e^{2\pi i n/N}\Phi_{n,m}$&$\Psi_{n,m+1}$ \\
\hline
\end{tabular}
 \caption{Action of the operators on the $Q-$ and $P-$ basis states
  \label{t}}
\end{table}
  
\begin{equation}\label{ft}
\Psi_{nm}(q,p) =\Sigma_{r,s=0}^{N-1} ~~e^{-2\pi i (nr+ms)/N} \Phi_{sr}(q,p).
\end{equation}
This is the analog of Fourier transform connecting the position space wave function to the momentum space 
wave function. Note however that the Fourier expansion  with respect to the first (second) index of  $\Psi_{nm}$
relates it to the second (first) index of $\Phi_{sr}$.

In our formalism the infinite dimensional prequantum Hilbert space results in a finite dimensional physical Hilbert space
as a consequence of the the equivalences of the basis states,
\begin{eqnarray}\nonumber
\psi'_{n,m}(q,p) \equiv \psi'_{n,m'}(q,p) \equiv \psi'_{n+N,m}(q,p) \equiv \psi'_{n,m+N}(q,p).\label{eq1}
\end{eqnarray}
Analogous equivalences are valid in the P-basis.

We may use any one phase space wave function out of the equivalent set for calculations. For instance, we 
may use $\psi'_{n,0}(q,p), n=0,1,2,\ldots ,(N-1)$. This is an orthonormal set with the inner product (\ref{tip1}).
All relevant operators are built using $exp(2\pi i\overleftarrow{Q}/b)$ and $exp(-2\pi i\overleftarrow{P}/a)$.
Because of the equivalences their action on these states are to be modified as follows:

\begin{equation}\label{tobs21}
e^{-2\pi i\overleftarrow{P}/a} \psi'_{n,0}(q,p) =\psi'_{(n+1)~(mod ~N), 0}(q,p).
\end{equation}

With all the equivalences, the physical Hilbert space is very simple, as is to be expected. It has dimension $N$.
In the Q-basis $exp(2\pi i\overleftarrow{Q}/a)$ has eigenvalues $e^{2\pi i n/N}, n=0,1,2,\ldots (N-1)$ and the operator $exp(-2\pi i\overleftarrow{P}/b) $ acts as the cyclic shift operator. Similar is the situation in the P-basis, with $q$ and $-p$
interchanged.This can be interpreted in terms of $N$ atoms spaced uniformly on a circle of circumference $b$. These correspond to the (periodic) delta function basis of  \cite{hb}.

\section{Discussion}\label{d}
In this paper we have advocated a combined use of the left and right invariant vector fields of the Heisenberg 
group for handling quantization of $R^2$ and torus phase spaces. Our aim is to obtain explicit canonical bases, 
operators and their actions. We  obtained equivalences of basis states in the prequantum Hilbert space,
which makes it equivalent to the standard Schr\"{o}dinger formalism. In case of torus we used the arguments of Dirac and of 
Wu and Yang for monopoles for obtaining the quantization of the phase space area in units of the Planck constant.
Of course this is equivalent to the cohomology arguments. We also demonstrated how the dimension of the physical
Hilbert space matches with the area of the phase space.

Our approach is applicable \cite{sg} to  any manifold with a symplectic structure, whether of unbounded or of a finite 
phase space volume. We obtain explicit canonical bases and operators using Darboux theorem.
In a subsequent paper \cite{s} we apply our techniques
to the 2-sphere $S^2$ as the classical phase space. This has direct connections to the Dirac's theory of  monopoles,
monopole harmonics, spin weighted spherical harmonics of Newman and Penrose. It is also related to  the  
Kirillov orbit method \cite{k} for $SU(2)$ group.

The extensive work  work on quantum cat map on a torus \cite{hb}  uses Wigner or Husimi distribution for
analyzing time evolution. 
It is interesting to use our wave functions to analyze this.


\end{document}